\documentclass[12pt]{iopart}
\usepackage{iopams}
\usepackage{cite}
\usepackage{pstricks}
\usepackage{epsfig}
\begin{document}
%----------------------------------------------------------------------------
%                              TITLE
%----------------------------------------------------------------------------
\title[Model C critical dynamics of  random anisotropy magnets]{
Model C critical dynamics of  random anisotropy magnets}
%----------------------------------------------------------------------------
%                              AUTHORS
%----------------------------------------------------------------------------

\author{M. Dudka$^{1,2}$, R. Folk$^{2}$,
 Yu.\ Holovatch$^{1,2}$ and G. Moser$^{3}$}

\address{$^{1}$
Institute for Condensed Matter Physics, National Acad. Sci. of Ukraine, UA-79011 Lviv, Ukraine}

\address{$^{2}$ Institut f\"ur Theoretische Physik, Johannes
Kepler Universit\"at Linz,  A--4040 Linz, Austria}

\address{$^{3}$
Institut f\"ur Physik und Biophysik, Universit\"at Salzburg, A--5020
Salzburg, Austria}

%----------------------------------------------------------------------------
%                             ABSTRACT
%----------------------------------------------------------------------------
\begin{abstract}
We study the relaxational critical dynamics of the three-dimensional
random anisotropy magnets with the non-conserved $n$-component order
parameter coupled to a conserved scalar density. In the random
anisotropy magnets the structural disorder is present in a form of
local quenched anisotropy axes of random orientation. When the
anisotropy axes are randomly distributed along the edges of the
$n$-dimensional hypercube, {\em asymptotical} dynamical critical
properties coincide with those of the random-site Ising model.
However structural disorder gives rise to considerable effects for
non-asymptotic critical dynamics. We investigate this phenomenon by
a field-theoretical renormalization group analysis in the two-loop
order. We study critical slowing down and obtain quantitative
estimates for the effective and asymptotic critical exponents of the
order parameter and scalar density. The results predict complex
scenarios for the effective critical exponent approaching an
asymptotic regime.
\end{abstract}
\pacs{05.50.+q, 05.70.Jk, 61.43.-j, 64.60.Ak, 64.60.Ht}
%\submitto{\JPA}

\eads{\mailto{maxdudka@icmp.lviv.ua}, \mailto{hol@icmp.lviv.ua},
\mailto{folk@tphys.uni-linz.ac.at}}

\maketitle

\section{Introduction}

In this paper, we address the peculiarities of criticality under an
influence of the  random anisotropy of structure. To be more
specific, given a reference system  is a 3d magnet with
$n$-component order parameter which below the second order phase
transition point $T_c$ characterizes a ferromagnetic state, what
will be the impact of random
anisotropy\cite{Harris73,Cochrane78,Dudka05} on the {\em critical
dynamics}\cite{Halperin77,Folk06} of this transition? It appears,
that contrary to the general believe that even weak random
anisotropy destroys ferromagnetic long-range order at $d=3$, this is
true only for the isotropic random axis distribution \cite{imryma}.
Therefore, we will study a particular case, when the second order
phase transition survives  and, moreover, it remains in the
random-Ising universality class \cite{Mukamel82,RAM_cub} for {\em
any} $n$. A particular feature of 3d systems which belong to the
random-Ising universality class is that their heat capacity does not
diverge at $T_c$ (it is the isothermal magnetic susceptibility which
manifests singularity) \cite{review}. Again, general arguments
state\cite{Krey76,Lawrie84} that for such systems the relaxational
critical dynamics of the non-conserved order parameter coupled to a
conserved density,  model C dynamics, degenerates to purely
relaxation model without any couplings to conserved densities (model
A). Nevertheless, this statement is true only in the
asymptotics\cite{Dudka06a,Dudka06b} (i.e. at $T_c$, which in fact is
never reached in experiments or in simulations). As we will show in
the paper, common influence of two different factors: randomness of
structure and coupling of dynamical modes leads to a rich effective
critical behavior which possesses many new unexpected features.

Dynamical properties of a system near the critical point are
determined by the behavior of its slow densities. In addition to the
order parameter density $\varphi$ these are the conserved densities.
Here, we consider the case of one conserved density $m$. For the
description of critical dynamics the characteristic time scales for
the order parameter, $t_{\varphi}$, and for the conserved density,
$t_{m}$, are introduced. Approaching the critical point, where the
correlation length $\xi$ is infinite, they are growing accordingly
to  the scaling laws
 \begin{eqnarray}\label{expon}
t_{\varphi}\sim\xi^z,\\ t_{m}\sim\xi^{z_{m}} \, .
\end{eqnarray}
These power laws define the dynamical critical exponents of the
order parameter, $z$,  and of the conserved densities, $z_{m}$. The
conserved density dynamical exponents may be different from that of
the order parameter.

The simplest dynamical model taking into account conserved densities
is model C,\cite{Halperin77,Halperin74} which contains a static
coupling between non-conserved $n$-dimensional order parameter
$\varphi$ and scalar conserved density $m$. Being quite simple, the
model can be applied  to the description of different physical
systems. In particular,  in a  lattice model of intermetallic
alloys\cite{Corentsveig97} the non-conserved order parameter
corresponds to differences in the concentration of atoms of certain
kind between the odd and even sublattices.  It is coupled to a
conserved quantity -- the concentration of atoms of this kind in the
full system. In the supercooled liquids the fraction of locally
favored structures is non-conserved ``bond order parameter", coupled
to the conserved density of a liquid \cite{Tanaka99}. Systems
containing annealed impurities with long relaxational
times\cite{Grinstein77} manifest certain similarity with the model C
as well.

Dynamical properties of a model with coupling to a conserved density
were less studied numerically than those for model without any
coupling to secondary densities. It may be the consequence of the
complexity of the numerical algorithms, which turn out  to be much
slower than for the simpler model. Simulations were performed for an
Ising antiferromagnet with conserved  full magnetization and
non-conserved staggered magnetization  (i.e. the order
parameter)\cite{Sen98} and also for an Ising magnet with conserved
energy \cite{Stauffer}.

Theoretical analysis of model C critical dynamics were performed by
means of the field-theoretical renormalization group. Critical
dynamical behavior of model C in different regions of $d-n$ plane
was analyzed by $\varepsilon=4-d$ expansion in first order in
$\varepsilon$ \cite{Halperin74}. The results lead to speculations
about the existence of an anomalous region for $2<n<4$, where the
order parameter is much faster than the conserved density and
dynamic scaling is questionable. Recent two-loop
calculation\cite{Folk03,Folk04} corrected the results of Ref.
\cite{Brezin75} and showed an absence of the anomalous region
$2<n<4$.

For  the 3d model C with  order parameter dimension $n=1$, the
conserved density lead to the "strong" scaling:\cite{Folk03,Folk04}
the dynamical exponents $z$ and $z_m$ coincide and are equal
$2+\alpha/\nu$, where $\alpha$ and $\nu$ are the specific heat and
the correlation length critical exponents, correspondingly. For the
Ising system ($n=1$) the specific heat diverges and  $\alpha>0$.
While for a system with $\alpha<0$, that is for the physically
interesting cases  $n=2,3$, the scalar density decouples from the
order parameter density in the asymptotic region. It means that for
such values of $n$ the order parameter  scales with the same
dynamical critical exponent $z$ as in the model A  and the dynamical
exponent of the scalar density  is equal to $z_m=2$. The importance
of the sign of $\alpha$ was  already mentioned in
Ref.~\cite{Halperin74}.

A rich critical dynamical behavior has already been observed in
system with structural disorder
\cite{Grinstein77,Prudnikov92,Oerding95,Janssen95,Blavatska05}.
Interest in this case is increased by the fact that real materials
are always characterized by some imperfection of their structure.
Obviously, that  models describing their properties should contain
terms connected with structural disorder of certain type. For the
static behavior of a system with quenched energy coupled disorder
(e.g. dilution), the Harris criterion\cite{Harris74} states that
disorder does not lead to a new static universality class if the
heat capacity of the pure system does not diverge, that is
$\alpha<0$. In appears that in diluted systems  $\alpha<0$ is always
the case (see Ref. \cite{review}). The conclusion about influence of
coupling between order parameter and secondary density works also in
this case. The presence of a secondary density does not affect the
dynamical critical properties in the asymptotics\cite{Krey76}: order
parameter dynamics  is the same as in an appropriate model A, and
$z_m=2$.  Nevertheless, as we noted at the beginning, the coupling
between the order parameter and the secondary density considerably
influences the non-asymptotic critical
behavior\cite{Dudka06a,Dudka06b}.

We are interested in the critical dynamics of a systems with
structural disorder of another type, namely,  random anisotropy
magnets. Their properties are described by the random anisotropy
model (RAM) introduced in Ref. \cite{Harris73}. In this spin lattice
model each spin is subjected to a local anisotropy of random
orientation, which essentially is described by a vector and
therefore is defined only for $n>1$. The Hamiltonian
reads:\cite{Harris73}
\begin{equation}
{\mathcal H} =  - \sum_{{\bf R},{\bf R'}} J_{{\bf R},{\bf R'}}
\vec{S}_{\bf R} \vec{S}_{\bf R'} -\bar{D}\sum_{{\bf R}} (\hat
{x}_{\bf R}\vec{S}_{\bf R})^{2},
 \label{origham}
\end{equation}
where,  $\vec{S}=({S^1},...,{S^n})$, are $n$-component vectors
located on the sites ${\bf R}$  of a  $d$-dimensional cubic lattice,
\mbox{$\bar{D}>0$} is an anisotropy constant, $\hat {x}$ is a random
unit vector pointing in direction of the local anisotropy axis. The
short-range interaction $J_{{\bf R},{\bf R'}}$ is assumed to be
ferromagnetic.

The static critical behavior of RAM was analyzed  by many
theoretical and numerical investigations which could be compared
with the critical properties of random anisotropy magnets found in
experiments (for recent review see Ref. \cite{Dudka05}). The results
of this analysis bring about that random anisotropy magnets do not
show a second order phase transition for an isotropic random axis
distribution. However they  possibly undergo a second-order phase
transition for an anisotropic distribution (for references see
reviews Refs. \cite{Cochrane78,Dudka05}). Renormalization group
studies of the
asymptotic\cite{Aharony75,RAM,Mukamel82,RAM_cub,new_Ital} and
non-asymptotic properties\cite{Dudka05} of RAM corroborated such a
conclusion. For example,  the  RAM with random axes distributed due
to the so-called cubic distribution  was shown within two-loop
approximation to undergo a second order phase transition governed by
the random Ising critical exponents,\cite{RAM_cub,Dudka05} as first
suggested in Ref. \cite{Mukamel82}. Recently this result found its
confirmation in a five-loop RG study \cite{new_Ital}. The cubic
distribution allows  $\hat x$ to point only along one of the $2n$
directions of the axes $\hat k_i$ of a (hyper)cubic
lattice:\cite{Aharony75}
\begin{equation}\label{cubdist}
p(\hat x) = \frac{1}{2n}\sum_{i=1}^n\left[\delta^{(n)}(\hat x-\hat
k_i)+\delta^{(n)}(\hat x+\hat k_i)\right],
\end{equation}
where $\delta(y)$ are Kronecker's deltas.

Contrary to the static critical behavior of random anisotropy
magnets their  dynamics was less investigated. Only dynamical models
for systems with isotropic distribution were briefly discussed in
Refs. \cite{Ma78,DeDominicis78}. The critical dynamics was discussed
within model A, Ref. \cite{Krey77}, and the dynamical exponents were
calculated. However, it does not give a comprehensive quantitative
description since it is (i) restricted to the isotropic distribution
of the random axis and (ii) it is performed only within the first
non-trivial order of $\varepsilon=4-d$ expansion.

The model A critical dynamics  of RAM with cubic random axis
distribution was analyzed within two-loop approximation in Ref.
\cite{Dudka06} Although the asymptotic dynamical properties found
coincide with those of the random-site Ising model, the
non-asymptotic behavior is strongly influenced by the presence of
random anisotropy \cite{Dudka06}.

Beside the slow order parameter an additional slow conserved
densities might be present, for instance the energy density.
Therefore considering the non-asymptotic dynamical behavior of the
RAM an extension to model C is of interest. Indeed, there exist
magnets where the distribution of the local random axes is
anisotropic (e.g. the rare earth compounds, see Ref.
\cite{Dudka05}).

The structure of the paper is as follows: Section \ref{II} presents
the equations defining the dynamical model and its Lagrangian, the
renormalization is performed is Section \ref{III}, there the
asymptotic and effective dynamical critical exponents are defined.
In Section \ref{IV} we give the expressions for the field-theoretic
functions in two-loop order and the resulting non-asymptotic
behavior is discussed. Section \ref{V} summarizes our study. Details
of the perturbation expansion are presented in the appendix.

\section{Model equations \label{II}}
Here we consider the dynamical model for random anisotropy systems
described by (\ref{origham}) with random axis distribution
(\ref{cubdist}). The structure of the equations of motion for
$n$-component order parameter $\vec{\varphi}_0$ and secondary
density\cite{Halperin74,Halperin77} $m_0$  is not changed by
presence of random anisotropy
\begin{eqnarray}\label{eq_mov2}
\frac{\partial {\varphi}_{i,0}}{\partial
t}&=&-\mathring{\Gamma}\frac{\partial {\mathcal H}}{\partial
{\varphi}_{i,0}}+{\theta}_{{\varphi}_{i}}, \qquad i=1\ldots n,\\
\label{eq_mov2v}
 \frac{\partial {m}_0}{\partial
t}&=&\mathring{\lambda}\nabla^2\frac{\partial {\mathcal H}}{\partial
{m}_0}+{\theta_{{m}}} \, .
\end{eqnarray}
The order parameter relaxes and conserved density diffuses with the
kinetic coefficients $\mathring{\Gamma}$, $\mathring{\lambda}$
correspondingly. The stochastic forces ${\theta}_{\varphi_i}$,
${\theta}_{m}$ obey the Einstein relations:
\begin{eqnarray}\label{1}
<\!{\theta}_{\varphi_i}(x,t){\theta}_{\varphi_j}(x',t')\!>\!\!&=&2\mathring{\Gamma}\delta(x-x')\delta(t-t')\delta_{ij},
\\ \label{2}
 <\!{\theta}_{m}(x,t){\theta}_{{m}}(x',t')\!\!>\!&=&\!\!{-}2\mathring{\lambda}\!\nabla^2\!\delta(x{-}x')\delta(t{-}t')\delta_{ij}
\, .
\end{eqnarray}
The disorder-dependent equilibrium effective static functional
$\mathcal H$ describing behavior of system in the equilibrium reads:
\begin{equation}
\label{effram} \hspace{-5em} {\cal H}{=} \int d{\bf R}
\left\{{1\over 2} \left[|\nabla
\vec{\varphi}_0|^2{+}\mathring{\tilde r}
|\vec{\varphi}_0|^2\right]\!{+} \frac{\mathring{\tilde v}}{4!}
|\vec{\varphi}_0|^4{-}\! D_0\!\left( \hat x
\vec{\varphi}_0\right)^2+ \frac{1}{2}{{m^2_0}}+
\frac{1}{2}\mathring{\gamma}
{{m}_0}|\vec{\varphi}_0|^2-\mathring{h}{{m}_0}\right\},
\end{equation}
where $D_0$ is an anisotropy constant proportional to $\bar{D}$ of
Eq. (\ref{origham}), $\mathring{\tilde r}$ and $\mathring{\tilde v}$
depend on $\bar{D}$ and the coupling of the usual $\phi^4$ model.

Integrating out the secondary density one reduces (\ref{effram}) to
usual Ginzburg-Landau-Wilson model with random anisotropy term and
new parameter $\mathring{v}$ and  $\mathring{r}$ connected to the
model parameters $\mathring{\tilde r},\mathring{\tilde v},
\mathring{\gamma}$ and $h$ via relations:
\begin{equation}
\mathring{r}=\mathring{\tilde r}+\mathring{\gamma} \mathring{h},
\qquad \mathring{v}=\mathring{\tilde v}-3{\mathring{\gamma}^2}
\end{equation}

We  study the critical  dynamics by applying the
Bausch-Janssen-Wagner approach\cite{Bausch76} of dynamical
field-theoretical renormalization group (RG). In this approach, the
critical behavior is studied on the basis of long-distance and
long-time properties of the  Lagrangian incorporating features of
dynamical equations of the model. The model defined by expressions
(\ref{eq_mov2})-(\ref{effram}) within Bausch-Janssen-Wagner
formulation\cite{Bausch76} turns out to be described by an
unrenormalized Lagrangian:
\begin{eqnarray}\label{L}
{\mathcal L}&=&\int d {\bf R} dt \Bigg\{
{-}\mathring{\Gamma}\sum_{i=1}^n{\tilde{\varphi}_{0,i}}{\tilde{\varphi}_{0,i}}{+}
\sum_{i=1}^n{\tilde{\varphi}_{0,i}} \left({\frac{\partial}{\partial
t}} {+}\mathring{\Gamma}(\mathring{\tilde\mu}{-}
\nabla^2)\right)\varphi_{0,i}+ \nonumber\\  && \mathring{\lambda}
\tilde{m}_0\nabla^2\tilde{m}_0
+\tilde{m}_0\left({\frac{\partial}{\partial
t}}-\mathring{\lambda}\nabla^2\right)m_0 {+} \sum_{i}
\frac{1}{3!}\mathring{\Gamma}\mathring{\tilde
v}\tilde{\varphi}_{0,i} {\varphi}_{0,i}\sum_j{\varphi}_{0,j}
{\varphi}_{0,j}+ \nonumber\\ &&  \sum_{j}2\mathring{\Gamma}D_0 (\hat
x \tilde{\varphi}_{0,i}(t))(\hat x {\varphi}_{0,i}(t))+
\mathring{\Gamma}\mathring{\gamma} m_0\tilde{\varphi}_{0,i}
{\varphi}_{0,i}{-} \frac{1}{2}\mathring{\lambda}\mathring{\gamma}
\tilde{m}_0\nabla^2{\varphi}_{0,i} {\varphi}_{0,i}\Bigg\},
\end{eqnarray}
 with  auxiliary response fields
${\tilde\varphi}_i(t)$. There are two ways to average over the
disorder configurations for dynamics. The first way originates from
statics and consists in using the replica trick,\cite{Emery75} where
$N$ replicas of the system are introduced in order to facilitate
configurational averaging of the corresponding generating
functional. Finally  the limit $N\to 0$ has to be taken.

However we follow the second way proposed in
Ref.~\cite{DeDominicis78}. There it was shown that the replica trick
is not necessary if one takes just the average of the Lagrangian
with respect to the distribution of random variables. The Lagrangian
obtained in this way is described by the following expression:
\begin{eqnarray}\label{rLagrangian}
{\mathcal L}& =& \Bigg\{\int d{\bf R} dt
\sum_i{\tilde{\varphi}_{i,0}}\Bigg[
 \left({\frac{\partial}{\partial t}}{+}\mathring{\Gamma}(\mathring{\tilde \mu}{-}
\nabla^2)\right)\varphi_{i,0}{-}\mathring{\Gamma}{\tilde{\varphi}_{i}}
 {+}\frac{\mathring{\Gamma}\mathring{{\tilde v}}}{3!}{\varphi_{i,0}}\sum_j{\varphi}_{j,0}{\varphi_{j,0}}
 {+} \nonumber\\&&  \frac{\mathring{\Gamma}{\mathring{y}}}{3!}
 {\varphi^3_{i,0}}\Bigg]{+}
\mathring{\lambda}
\tilde{m}_0\nabla^2\tilde{m}_0+\tilde{m}_0\left({\frac{\partial}{\partial
t}}-\mathring{\lambda}\nabla^2\right)m_0+\mathring{\Gamma}\mathring{\gamma}
m_0\tilde{\varphi}_{i,0} {\varphi}_{i,0}{-} \nonumber\\ &&
\frac{1}{2}\mathring{\lambda}\mathring{\gamma}
\tilde{m}_0\nabla^2{\varphi}_{i,0} {\varphi}_{i,0}{+} \int
dt'\sum_{i}\tilde{\varphi}_{i,0}(t) {\varphi_{i,0}}(t)\Bigg[
\frac{\mathring{\Gamma}^2{\mathring{u}}}{3!}\sum_{j}\tilde{\varphi}_{j,0}(t')
{\varphi_{j,0}}(t'){+} \nonumber\\ &&
\frac{\mathring{\Gamma}^2{\mathring{w}}}{3!}\tilde{\varphi}_{i,0}(t')
{\varphi_{i,0}}(t')\Bigg]\Bigg\} .
\end{eqnarray}

In Eq. (\ref{rLagrangian}), the bare mass is
$\mathring{\tilde\mu}=\mathring{\tilde r}-D/n$, and bare couplings
are $\mathring{u}>0$, $\mathring{\tilde v}>0$, $\mathring{w}<0$.
Terms with couplings $\mathring{u}$ and $\mathring{w}$ are generated
by averaging over configurations and the values of $\mathring{u}$
and $\mathring{w}$ are connected to the moments of distribution
(\ref{cubdist}). Therefore the ratio of the two couplings has to be
$\mathring{w}/\mathring{u}=-n$. The $\mathring{y}$-term in
(\ref{rLagrangian}) does not result from the averaging procedure but
has to be included since it is generated in the perturbational
treatment. It can be of either sign.

\section{RG functions \label{III}}

We perform renormalization within minimal subtraction scheme
introducing renormalization factors $Z_{a_i}$, $a_i=\{\{\alpha\},
\{\delta\}\}$, leading to the renormalized parameters
$\{\alpha\}=\{u,v,w,y,\gamma, \Gamma, \lambda \}$ and renormalized
densities $\{\delta\}=\{\varphi,\tilde\varphi,m,\tilde m\}$. For the
specific heat we need also an additive renormalization
$A_{\varphi^2}$ which leads to the function
\begin{equation}
B_{\varphi^2}(u,\Delta)=\mu^{\varepsilon}Z^2_{\varphi^2}\mu\frac{d}{d
\mu}\left(Z^{-2}_{\varphi^2}\mu^{-\varepsilon}A_{\varphi^2}\right) ,
\end{equation}
with the scale parameter $\mu$ and factor $Z_{\varphi^2}$ that
renormalizes  the vertex with $\varphi^2$ insertion.
  From the $Z$-factors one obtains the
$\zeta$-functions describing the critical properties
\begin{eqnarray}\label{def_z}
\zeta_{a}(\{\alpha\})&=&-\frac{d\ln Z_{a}}{d \ln \mu} \, ,
\end{eqnarray}
Relations between the renormalization factors lead to corresponding
relations between the $\zeta$-functions. In consequence for the
description of the critical dynamics one needs only
$\zeta$-functions of the couplings,
 $\zeta_{u_i}$ ($u_i=\{u,v,w,y\}$ for $i=1,2,3,4$),
 the order parameter
$\zeta_{\varphi}$, the auxiliary field $\zeta_{\tilde\varphi}$,
$\varphi^2$-insertion $\zeta_{\varphi^2}$ and also function
$B_{\varphi^2}$. In particular, the $\zeta$-function of the time
scale ratio
\begin{equation}
W=\frac{\Gamma}{\lambda}
\end{equation}
introduced for the description of dynamic properties is related to
the above $\zeta$-functions:
\begin{eqnarray}\label{zWW}
\zeta_{W}&=&\frac{1}{2}\zeta_{\varphi}-\frac{1}{2}{\zeta_{\tilde\varphi}}-\gamma^2
B_{\varphi^2} .
\end{eqnarray}

The behavior of the model parameters under renormalization is
described by the flow equations
\begin{equation}\label{fl}
\ell\frac{d\{ \alpha\}}{d \ell}=\beta_{\{\alpha\}} \, .
\end{equation}
The $\beta$-functions for the static model parameters have the
following explicit form:
\begin{eqnarray}
\beta_{u_i}&=&u_i(\varepsilon+\zeta_{\varphi}+\zeta_{u_i}),\\
\beta_{\gamma}&=&\gamma(\frac{\varepsilon}{2}+\zeta_{\varphi^2}+\frac{\gamma^2}{2}B_{\varphi^2}).
\label{gam}
\end{eqnarray}

The dynamic $\beta$-function for the time scale ratio $W$  reads
\begin{eqnarray}\label{WW}
\beta_{W}&=&W\zeta_{W}=W(\frac{1}{2}\zeta_{\varphi}-\frac{1}{2}{\zeta_{\tilde\varphi}}-\gamma^2
B_{\varphi^2}).
\end{eqnarray}

The asymptotic critical behavior of the system is obtained from the
knowledge of the fixed points (FPs) of the flow equations
(\ref{fl}). A FP $\{\alpha^*\}=\{u^*,v^*,w^*,y^*,\gamma^*,W^*\}$ is
defined as simultaneous zero of the $\beta$-functions. The set of
equations for the static fourth order couplings decouple from the
other $\beta$-functions. Thus for each of the FPs of the static
forth order couplings $\{u_i^*\}$ one obtains two FP values of the
static coupling between the order parameter and the conserved
density $\gamma$:
 \begin{equation} {\gamma^*}^2{=}0
\quad {\rm and} \quad
{\gamma^*}^2{=}\frac{\varepsilon-2\zeta_{\varphi^2}(\{u^{\star}_i\})}{B_{\varphi^2}(\{u^{\star}_i\})}
{=}\frac{\alpha}{\nu B_{\varphi^2}(\{u^{\star}_i\})},
\end{equation}
where $\alpha$ and $\nu$ are the heat capacity and correlation
length critical exponent calculated at the corresponding FP
$\{u^*\}$. Inserting the obtained values for the static FPs into the
$\beta$-function (\ref{WW}) one finds the corresponding FP values of
the time scale ratio $W$.

The stable FP accessible from the initial conditions  corresponds to
the critical point of system. A FP is stable if all eigenvalues
$\omega_i$ of the stability matrix
${\partial\beta_{\alpha_i}}/{\partial {\alpha_j}}$ calculated at
this FP have  positive real parts. The values of $\omega_i$ indicate
also how fast the renormalized model parameters reach their fixed
point values.

From the structure of $\beta$-functions  we conclude, that the
stability of any FP with respect to the parameters $\gamma$ and $W$
is determined solely by the derivatives of the corresponding
$\beta$-functions:
\begin{equation}
\omega_{\gamma}=\frac{\partial\beta_{\gamma}}{\partial
{\gamma}},\qquad \omega_{W}=\frac{\partial\beta_{W}}{\partial {W}}
\, .
\end{equation}
Moreover using (\ref{gam}) we can write:
\begin{equation}
\omega_{\gamma}=-\frac{\varepsilon-2\zeta_{\varphi^2}(\{u_i\})}{2}+\frac{3}{2}\gamma^{2}B_{\varphi^2}(u,\Delta)
\, ,
\end{equation}
which at the FP $\{\alpha^*\}$ leads to:
\begin{eqnarray}
\left.\omega_{\gamma}\right|_{{\{\alpha\}}={\{\alpha^*\}}}&=&-\frac{\alpha}{2\nu}
\qquad {\rm for} \qquad  {\gamma^*}^2=0 \, ,\\
\left.\omega_{\gamma}\right|_{{\{\alpha\}}={\{\alpha^*\}}}&=&\frac{\alpha}{\nu}
\qquad {\rm for} \qquad {\gamma^*}^2\neq 0 \, .
 \end{eqnarray}
Therefore, a stability with respect to parameter $\gamma$ is
determined by the sign of the specific heat exponent $\alpha$. For a
system with non-diverging heat capacity ($\alpha<0$) at  the
critical point, $\gamma^*=0$ is the stable FP.   Static results
report that the stable and accessible FP is of a random site Ising
type. In this case $\alpha<0$. This leads to the  conclusions that
in the asymptotic region the secondary density  decouples from the
order parameter.

The critical exponents are defined by the FP values of the
$\zeta$-functions. For instance, the asymptotic dynamical critical
exponent $z$  is expressed at the stable FP by:
\begin{equation}\label{zzz}
z=2+\zeta_{\Gamma}(\{\alpha^*\}),
\end{equation}
with
\begin{equation}\label{zg}
\zeta_{\Gamma}(\{\alpha\})=
\frac{1}{2}\zeta_{\varphi}(\{u_i\})-\frac{1}{2}\zeta_{\tilde\varphi}(\{\alpha\}).
\end{equation}
In similar way  the dynamical critical exponent $z_m$  for the
secondary density is defined by:
\begin{equation}
z_m=2+\zeta_m(\{u_i^*\},\gamma^*),
\end{equation}
where
\begin{equation}
\zeta_m(\{u_i\},\gamma)=\frac{1}{2}{\gamma}^2B_{\varphi^2}(\{u_i\}).
\end{equation}
While their effective counterparts in the non asymptotic region are
defined by the solution of flow equations (\ref{fl}) as
\begin{equation}\label{zzz_eff}
z^{\rm eff}=2+\zeta_{\Gamma}(\{u_i(\ell)\},\gamma(\ell),W(\ell)),
\end{equation}
\begin{equation}
z_{m}^{\rm eff}=2+\zeta_m(\{u_i(\ell)\},\gamma^2(\ell)).
\end{equation}
In the limit $\ell\to 0$ the effective  exponents reach their
asymptotic values. In the next section we analyze the possible
scenarios of effective dynamical behavior as well as check the
approach to the asymptotical regime.

\section{Results \label{IV}}
\subsection{Asymptotic properties}
The static two-loop RG functions of RAM with cubic random axis
distribution in the minimal substraction scheme agree with the
results obtained in Ref. \cite{Dudka05} using the replica trick and
read:

\begin{eqnarray}\label{betacub1}
\beta_u\!&=&\!-\varepsilon u{+}\frac {4}{3}{u}^{2}{+}\frac{n+2}{3}vu
{+}\frac{2}{3}uw{+}yu{+}\frac{1}{3}wv{-} \frac{7}{6}{u}^{3}{-}{\frac
{11\left (n+2\right)}{18}}v{u}^{2} \nonumber\\
&& {-}{\frac {5\left (n+2\right )}{36}}{v}^{2}u {- }{\frac
{11}{9}u^{2}w}{-} {\frac {5}{18}}u{w}^{2}{-}{\frac
{11}{6}}{u}^{2}y-{\frac {5}{12}}{y}^{2}u {-}\nonumber\\
&&\frac{3}{2}vuw-\frac{5}{6}wyu-\frac{5}{6}vuy
-\frac{1}{9}{v}^{2}w-\frac{1}{9}{w}^{2}v,
\end{eqnarray}

\begin{eqnarray}\label{betacub2}
\beta_v\!&=&\!-\varepsilon v{+}\frac
{n+8}{6}{v}^{2}{+}2vu{+}\frac{2}{3}wv{+}yv
{-}\frac{3n+14}{12}{v}^{3}{-} {\frac {11n{+}58}{18}}{v}^{2}u
{-}{\frac {41}{18}}v{u}^{2} {-}\nonumber\\ &&  {\frac
{5}{18}}{w}^{2}v {-}\frac{5}{6}vwy{-} {\frac
{31}{18}}{v}^{2}w{-}{\frac {11}{6}}{v}^{2}y {-}{\frac
{5}{12}}{y}^{2}v{-}{\frac {17}{6}}vuy{-}{\frac {17}{9}}wvu,
\end{eqnarray}

\begin{eqnarray}\label{betacub3}
\beta_w\!&=&\!-\varepsilon w{+}\frac
{4}{3}{w}^{2}{+}2wu{+}\frac{2}{3}wv{+}yw {-} \frac{7}{6}{w}^{3}{-}
{\frac
{29}{9}}{w}^{2}u{-}{\frac {41}{18}}w{u}^{2} {-} {\frac {31}{18}}{w}^{2}v{-}
 \nonumber\\
&&{\frac {n{+}10}{36}}{v}^{2}w {-}{\frac {11}{6}}{w}^{2}y {-}{\frac
{5}{12}}{y}^{2}w{-}{\frac {17}{6}}wuy{-}{\frac
{5n+34}{18}}wvu{-}\frac{5}{6}vwy,
\end{eqnarray}

\begin{eqnarray}\label{betacub4}
\beta_y&=&-\varepsilon
y+\frac{3}{2}{y}^{2}+2yu+2yv+2wy+\frac{4}{3}wv{-} {\frac
{17}{12}}{y}^{3}- {\frac {41}{18}}{u}^{2}y{-} \nonumber\\
&& {\frac {23}{6}}{y}^{2}u{-}{\frac {23}{6}}{y}^{2}v {-}{\frac
{23}{6}}{y}^{2}w{-}{\frac {5n+82}{36}}{v}^{2}y{-}{ \frac
{41}{18}}{w}^{2}y{-} {2}{w}^{2}v{-} \nonumber\\&& \frac
{n+18}{9}{v}^{2}w{-} {\frac {41}{6}}vwy{ -}{\frac
{41}{9}}wyu{-}{\frac {5n+82}{18}}vuy{-}\frac{8}{3}vuw,
\end{eqnarray}

\begin{equation}\label{gammacub1}
\zeta_\varphi={\frac {1}{36}}{u}^{2}{+}\frac{{y}^{2}}{24}{+}{\frac
{1}{36}}{w}^{2} {+}{\frac
{n+2}{72}}{v}^{2}{+}\frac{yu}{12}{+}\frac{wv}{12}{+}\frac{yv}{12}{+}\frac{n+2}{36}
vu{+}\frac {wu}{18}{+}\frac{wy}{12},
\end{equation}

\begin{eqnarray}\label{gammacub2}
{\zeta}_{\varphi^2}&{=}&\frac{1}{3}u{+} \frac {n+2}{6}v{+}\frac
{1}{3}w+\frac {y}{2}{-}\frac {5}{12}{u}^{2}{-} \frac
5{n+2}{12}{v}^{2}{-}\frac{5}{6}{w}^{2}{-}\frac {5}{4}{y}^{2}
{-}\nonumber\\ &&5\frac {n+2}{6}vu{-}\frac
{5}{3}wu{-}5\frac{yu}{2}{-}5\frac{wv}{2}{-}5\frac{yv}{2}{-}5\frac{wy}{2}
\, .
\end{eqnarray}
Here, $u,v,w,y$ stand for the renormalized couplings.

Given the expression  for the function ${\zeta}_{\varphi^2}$,  Eq.
(\ref{gammacub2}), the function $\beta_{\gamma}$ can be constructed
via Eq.  (\ref{gam}) and the two-loop expression
$B_{\varphi^2}=n/2$.

 In order
to discuss the dynamical FPs it turns out to be useful to introduce
the parameter $\rho=W/(1+W)$ which maps $W$ and its FPs
 into a finite region of the parameter
space $\rho$. Then instead of the flow equation for $W$ the flow
equation for $\rho$ arises in (\ref{fl}):
\begin{equation}\label{drho}
\ell\frac{d\rho}{d\ell}=\beta_{\rho}(\{u_i\},\gamma,\rho),
\end{equation}
where according to (\ref{WW})
\begin{equation}\label{brho}
\beta_{\rho}(\{u_i\},\gamma,\rho)=\rho(\rho-1)({\zeta_{\Gamma}}(\{u_i\},\gamma,\rho)-\gamma^2
B_{\varphi^2}(\{u_i\})).
\end{equation}

The function $\zeta_{\Gamma}$ in the above expression is obtained
from Eq (\ref{zg}) using the static function $\zeta_\varphi$
(\ref{gammacub1}) and the two loop result for the dynamic function
$\zeta_{\tilde\varphi}$ (calculated from Eq (\ref{zphitilde})). We
get the following two-loop expression for $\zeta_{\Gamma}$:
\begin{eqnarray}\label{resultat}
\zeta_{\Gamma}&=&-\frac{u+w}{3}+{{\gamma}}^{2}\rho+\frac{\left( 6\ln
\left( 4/3
 \right) -1 \right)}{24} \left( \frac{\left( n+2 \right)}{3} {v}^{2}+\frac{2}{3}vy+{
y}^{2} \right) +\nonumber\\&&\frac{1}{{36}} \left({5{u}^{2}+\left(
n+2 \right) uv+10uw+3uy+3vw+{5}{w}^{2}+3wy}\right)-
\nonumber\\&&\frac{\rho{\gamma}^{2}}{2} \Bigg( \left( \frac{\left(
n+2 \right)}{3}v +y \right)  \left( 1-3\ln \left( 4/3 \right)
\right)+ \nonumber\\&& \rho{{\gamma}}^{2} \left(
\frac{n}{2}-\rho-\frac{3 \left( n+2 \right)}{2} \ln  \left( 4/3
 \right) + \left( 1+\rho \right) \ln  \left( 1-{\rho}^{2} \right)
 \right) +u+w \Bigg) +\nonumber\\&&{{\gamma}}^{2}\rho \left( \frac{u+w}{6}
 \right)  \left( {\frac {{\rho}^{2}\ln  \left( \rho \right) }{1-\rho}}
+ \left( 3+\rho \right) \ln  \left( 1-\rho \right)  \right).
\end{eqnarray}
The two-loop result\cite{Folk04} for the pure model C is recovered
by setting in (\ref{resultat}) the couplings $u,w,y$ equal to zero.
While setting  $\gamma=0$ in (\ref{resultat}) the result for model A
with random anisotropy\cite{Dudka06} is recovered. The
$\gamma^2u,\gamma^2w,\gamma^2y$-terms represent the intrinsic
contribution of model C for random anisotropy magnets.

 There are two different ways to proceed with the numerical analysis
of the perturbative expansions for the RG functions (\ref{betacub1})
- (\ref{betacub4}), (\ref{resultat}). The first one is an
$\varepsilon$-expansion \cite{Wilson72} whereas the second one is
the so-called fixed-dimension approach \cite{Schloms}. In the frames
of the latter approach, one fixed $\varepsilon$ and solves the
non-linear FP equations directly at the space dimension of interest
(i.e. at $\varepsilon=1$ in our $d=3$ case). Whilst in many problems
these two ways serve as complementing ones, it appears that for
certain cases only one of them, namely the fixed-$d$ approach leads
to the quantitative description. Indeed, as it is well known by now,
the $\varepsilon$-expansion turns into the
$\sqrt{\varepsilon}$-expansion for the random-site Ising model and
no reliable numerical estimates can be obtained on its basis (see
\cite{review} and references therein). As one will see below, the
random-site Ising model behavior emerges in our problem as well,
therefore we proceed within the fixed-$d$ approach.

The series for  RG functions are known to diverge. Therefore to
obtain reliable results on their basis  we apply the Pad\'e-Borel
resummation procedure\cite{Baker78} to the static functions. It is
performed in following way: we construct the Borel image of the
resolvent series\cite{Watson74} of the initial RG function $f$:
\begin{eqnarray*}
f=&\sum_{0\le i+j+l+k\le2}a_{i,j,k,l}(ut)^i(vt)^j(wt)^k(yt)^l&\to\\
&\sum_{0\le i+j+l+k\le2}\frac{a_{i,j,k,l}u^iv^jw^ky^l
t^{i+j+k+l}}{\Gamma(i+j+1)},&
 \end{eqnarray*}
  where $f$ stands for one of the static RG functions $\beta_{u_i}$,
$\beta_{\gamma}/\gamma-\gamma^2n/4$, $a_{i,j,k,l}$ are the
corresponding expansion coefficients given by Eqs.
(\ref{betacub1})--(\ref{gammacub2}), and $\Gamma(i+j+1)$ is Euler's
gamma function. Then, the Borel image is extrapolated by a rational
Pad\'e approximant\cite{Baker81} $[K/L](t)$. Within two-loop
approximation we use the diagonal approximant with linear
denominator [1/1]. As it is known in the Pad\'e analysis, the
diagonal approximants ensure the best convergence of the results
\cite{Baker81}. The resummed function is then calculated by an
inverse Borel transform of this approximant:
\begin{equation}\label{res}
f^{res}=\int_0^{\infty}{\rm d}t \exp(-t)[1/1](t).
\end{equation}
As far as the above procedure enables one to restore correct static
RG flow (as sketched below) we do not further resum the dynamic RG
function $\beta_W$.

The analysis of the static functions  $\beta_{\{u_i\}}$ at fixed
dimension $d=3$ brings about an existence of 16 FPs
\cite{Dudka05,new_Ital}. Only ten of these FPs are situated in the
region of physical interest $u>0$, $v>0$, $w<0$. Corresponding
values of FP coordinates can be found in Ref. \cite{Dudka05}.

\begin{table}
\caption {\label{tab1} Two-loop values for the dynamical FPs of
random anisotropy magnets with $n=2$ (model C).}
   \begin{tabular}{|c|c|c|c|c|c|}
\hline FP &
$\gamma^*$&$\rho^*$&$\omega_{\gamma}$&$\omega_{\rho}$&$z$\\
   \hline
\hline  {\bf I}&0&$0\le\rho^*\le1$ &0& 0&2\\
 {\bf I$^{\prime}$} &1&0&1& -1 &2\\ {\bf I$_C$}&1&0.6106&1& 0.745 &3\\
  {\bf I$^{\prime}_1$}&1&1&1& -$\infty$ &$\infty$\\
 \hline
 {\bf II}&0&0&0.0387& 0.0526 &2.0526\\
 {\bf II$_1$}&0&1&0.0387&-0.0526  &2.0526\\
 \hline
 {\bf III}&0&0&-0.1686&-0.1850 &1.8150 \\
 {\bf III$_1$}&0&1&-0.1686& 0.1850 &1.8150\\
 {\bf III$_C$}&.5806&0&0.3371& -0.5222 &1.8150\\
 {\bf III$^{\prime}_1$}&.5806&1&0.3371& $\infty$ &$-\infty$\\
 \hline
 {\bf V} &0 &0 &-0.0525&0.0523&2.0523\\
 {\bf V$_1$} &0 &1 &-0.0525&-0.0523&2.0523\\
   {\bf V$^{\prime}$}&.3240&0&0.1050 & -0.0527&2.0523\\
   {\bf V${_C}$}&.3240&0.5241&0.1050 & 0.0277&2.1050\\
   {\bf V$^{\prime}_1$}&.3240&1&0.1050 & -$\infty$&$\infty$\\
 \hline
{\bf VI} &0 &0 &-0.0049&-0.0417&2.0107\\
 {\bf VI$_1$} &0 &1&-0.0049&0.0417&2.0107\\
 {\bf VI$^{\prime}$} &0.0986 &0 &0.0097&0.00095&2.0107\\
{\bf VI$^{\prime}_1$} &0.0986 &1 &0.0097&$\infty$&-$\infty$\\
\hline
 {\bf VIII} &0 &0 &-0.0525 &0.1569&2.1569\\
  {\bf VIII$_1$} &0 &1 &-0.0525 &-0.1569&2.1569\\
 {\bf VIII$^{\prime}$} &0.3240 &0&0.1050 &0.0519&2.1569\\
  {\bf VIII$^{\prime}_1$} &0.3240 &1 &0.1050 &-$\infty$&$\infty$\\
  \hline {\bf X}&0 &0
&-0.0525&0.0523&2.0523\\ {\bf X$_1$}&0 &1 &-0.0525&-0.0523&2.0523\\
{\bf X$^{\prime}$}&.3240&0&0.1050 & -0.0527&2.0523\\ {\bf
X${_C}$}&.3240&0.5241&0.1050 & 0.0277&2.1050\\ {\bf X$^{\prime}_1$}
&.3240&1&0.1050 &
-$\infty$&$\infty$\\
 \hline {\bf XV}&0 &0 &0.0018&0.1388&2.1388\\ {\bf XV$_1$}&0 &1
&0.0018&-0.1388&2.1388\\
  \hline
  \end{tabular}
  \end{table}

\begin{table}[htp]
\caption {\label{tab2} Two-loop values for the dynamical FPs of
random anisotropy magnets with $n=3$ (model C).}
   \begin{tabular}{|c|c|c|c|c|c|}
\hline FP &
$\gamma^*$&$\rho^*$&$\omega_{\gamma}$&$\omega_{\rho}$&$z$\\
   \hline
\hline  {\bf I}&0&$0\le\rho^*\le 1$&0& 0&2\\
    {\bf I$^{\prime}$}&0.8165&0&1& -1 &2\\
  {\bf I$_C$}&0.8165&0.7993&1&0.5218 &3\\
   {\bf I$^{\prime}_1$}&0.8165&1&1& -$\infty$
&$\infty$\\
 \hline
 {\bf II}&0&0&0.1109& 0.0506 &2.0506\\
{\bf II$_1$}&0&1&0.1109&-0.0506  &2.0506\\ \hline {\bf
III}&0&0&-0.1686&-0.1850 &1.8150\\
 {\bf III$_1$}&0&1&-0.1686& 0.1850 &1.8150\\
{\bf III$^{\prime}$}&0.4741&0&.3371& -0.5222 &1.8150\\ {\bf
III$^{\prime}_1$}&0.4741&1&0.3371& $\infty$ &$-\infty$\\
 \hline
 {\bf V} &0 &0 &-0.0525&0.0523&2.0523\\
{\bf VI$_1$} &0 &1 &-0.0525&-0.0523&2.0523\\
 {\bf VI$^{\prime}$}
 &0.2646&0&0.1050 & -0.0527&2.0523\\
 {\bf VI$_C$}&0.2646&0.7617&0.1050 & 0.0157&2.1050\\
 {\bf VI$^{\prime}_1$}&0.2646&1&0.1050 & -$\infty$&$\infty$\\
\hline {\bf VI} &0 &0 &-0.0162&-0.0401&1.9599\\
 {\bf VI$_1$} &0 &1 &-0.0162&0.0401&1.9599\\
 {\bf VI$^{\prime}$} &0.1467 &0 &0.0323&-0.0724&1.9599\\
 {\bf VI$^{\prime}_1$} &0.1467 &1  &0.0323&$\infty$&-$\infty$\\
 \hline
 {\bf VIII} &0 &0 &0.1051 &0.0425&2.0425\\
{\bf VIII$_1$} &0 &1 &0.1051 &-0.0425&2.0425\\
 \hline
 {\bf IX}&0 &0 &-0.0161&-0.0384&1.9616\\
{\bf IX$_1$}  &0 &1 &-0.0161&0.0384&1.9616\\
 {\bf IX$^{\prime}$} &0.1466&0&0.0322 & -0.0707&1.9616\\
 {\bf IX$^{\prime}_1$}   &0.1466&1&0.0322 & $\infty$&-$\infty$\\
 \hline
 {\bf X}&0 &0 &-0.0525&0.0523&2.0523\\
 {\bf X$_1$}&0 &1 &-0.0525&-0.0523&2.0523\\
 {\bf X$^{\prime}$}&0.2646&0&0.1050 & -0.0527&2.0523\\
{\bf X$_C$} &0.2646&0.7617&0.1050 & 0.0157&2.1050\\
 {\bf X$^{\prime}_1$}&0.2646&1&0.1050 & -$\infty$&$\infty$\\
{\bf XV}&0 &0 &0.0018&0.1388&2.1388\\
 {\bf XV$_1$}&0 &1 &0.0018&-0.1388&2.1388\\
 \hline
 \end{tabular}
\end{table}

For each static FP $\{u^*_i\}$ we obtain a set of dynamical FPs with
different $\gamma^*$ and $\rho^*$. The FPs obtained for $n=2,3$ are
listed in Table \ref{tab1} and Table \ref{tab2} correspondingly.
Stability exponents $\omega_{\gamma}$ and $\omega_{\rho}$ are given
in tables as well. Here we keep the numbering of FPs already used in
Refs. \cite{Aharony75,RAM_cub,new_Ital,Dudka05,Dudka06}.  It is
known from the statics  that FP XV governs the critical behavior of
RAM with cubic distribution. This FP is of the same origin as the FP
of random-site Ising model therefore all static critical exponents
coincide with those of the random-site Ising model. Since the
specific  heat exponent in this case is negative, the asymptotic
critical dynamics is described by model A. However the
non-asymptotic critical properties of random anisotropy magnets are
different from the random-site Ising magnets in
statics\cite{Dudka05} as well as in dynamics \cite{Dudka06}.
Moreover, for the model C considered here, the non-asymptotic
critical behavior differs considerably from that of the
corresponding model A as we will se below.

\begin{figure}[htbp]
\centerline{{\includegraphics[width=0.4\textwidth]{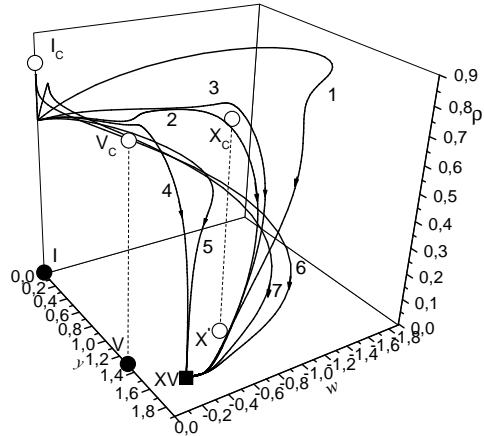}}}
\caption{\label{fig1}Projections of flows for $n=3$ in the
 subspace of couplings $w-y-\rho$. Open circles
represent projections of unstable FPs with non-zero $\gamma^*$.
Filled circles denote unstable FPs with $\gamma^*=0$. The filled
square shows the stable FP. See Section \ref{IVB} for a more
detailed description.}
\end{figure}

\begin{figure}[htbp]
\centerline{{\includegraphics[width=0.4\textwidth]{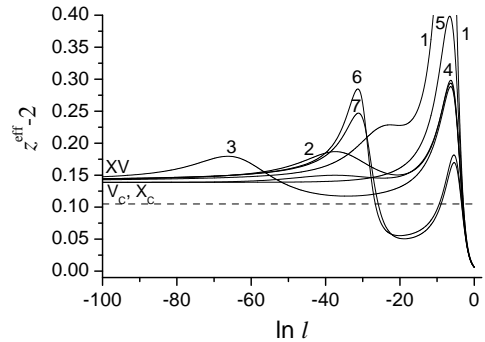}}}
\caption{\label{fig2} Dependence of the order parameter effective
dynamical critical exponent in the model C dynamics on the logarithm
of flow parameter. See text for full description.}
\end{figure}

\begin{figure}[htbp]
\centerline{{\includegraphics[width=0.4\textwidth]{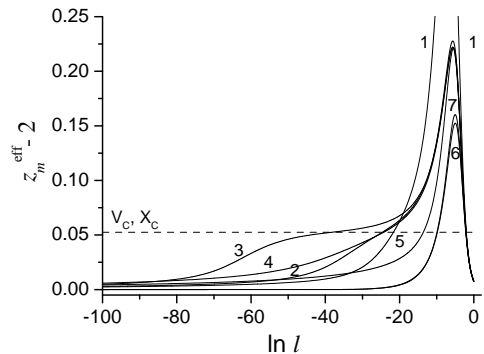}}}
\caption{\label{fig3}Dependence of the  conserved density effective
dynamical critical exponent in the model C dynamics on the logarithm
of flow parameter. See text for full description.}
\end{figure}

\subsection{Non-asymptotic properties}\label{IVB}
 The existence of such a large number of dynamical FPs makes non-asymptotic
critical behavior more complex as in model A. We present here
results for $n=3$. For $n=2$ the behavior is qualitatively similar.
Solving the flow equations for different initial conditions we
obtain different flows in the space of model parameters. The
projection of most characteristic flows  into the subspace
$w-y-\rho$ is presented in Fig. \ref{fig1}. The open circles
indicate genuine model C unstable FPs whereas filled circles
represent model A unstable FPs. The filled square denotes the stable
FP.

The initial conditions for the couplings $u(0), v(0), w(0), y(0)$
for the flows shown are the same as those in Refs.
\cite{Dudka05,Dudka06}. We choose $\gamma(0)=0.1$ and $\rho(0)=0.6$.
Many flows are affected by the two Ising FPs V$_C$ and X$_C$.
Inserting the solutions of the flow equations into the expressions
for dynamical exponents we obtain the effective exponents $z^{\rm
eff}$ and $z_{m}^{\rm eff}$. The dependence of $z^{\rm eff}$ on the
flow parameter $\ell$ corresponding to flows 1-7 is shown in
Fig.~\ref{fig2}. Similarly Fig~\ref{fig3} shows this dependence for
the effective exponent of the conserved density $z_{m}^{\rm eff}$.
Flow 3 is affected by both  FPs V$_C$ and X$_C$. Therefore the
effective exponents demonstrate a region with values which are close
to those for model C in the case of the Ising magnet (see curves 3
in Figs.~\ref{fig2} and \ref{fig3}). The asymptotic values
corresponding to the FPs  V$_C$ and X$_C$ are indicated by the
dashed line. They correspond to the values asymptotically obtained
in the pure model C with $n=1$, since the FPs V$_C$ and X$_C$ are of
the same origin, that  FP of pure model C. Curves 6 correspond to
flows near the pure FP II. Whereas curve 7 corresponds to the flow
near the cubic FP VIII.

\begin{figure}[htbp]
\centerline{{\includegraphics[width=0.4\textwidth]{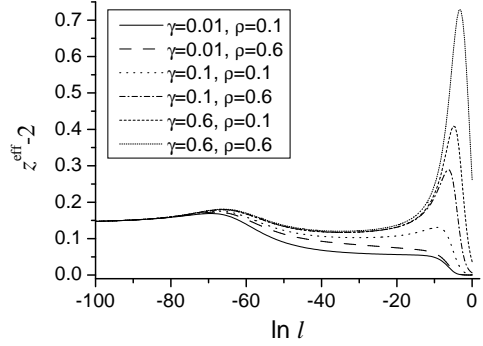}}}
\caption{\label{fig4}Dependence of $z^{\rm eff}$  on the logarithm
of flow parameter for different initial values $\gamma$ and $\rho$}
\end{figure}

The main difference of the behavior of the effective dynamical
exponent  $z^{\rm eff}$ in model C from that in model A is the
appearance of curves with several peaks. The value of the peak
appearing on the right-hand side depends on the initial condition
$\gamma(0)$ and $\rho(0)$. This is demonstrated in Fig.\ref{fig4}.

\begin{figure}[htbp]
\centerline{{\includegraphics[width=0.4\textwidth]{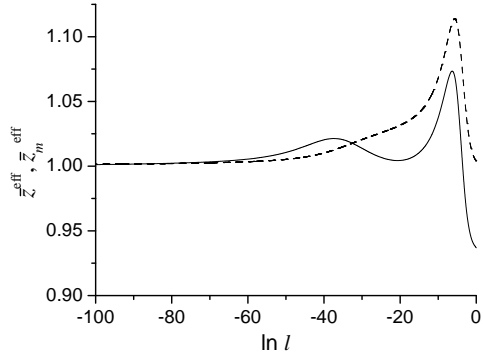}}}
\caption{\label{fig5} Normalized effective dynamical critical
exponents of order parameter and conserved density
$\overline{z}^{\rm eff}$ (solid line),  $\overline{z}_{m}^{\rm eff}$
(dashed line) correspondent to flow 2.}
\end{figure}

\begin{figure}[htbp]
\centerline{{\includegraphics[width=0.4\textwidth]{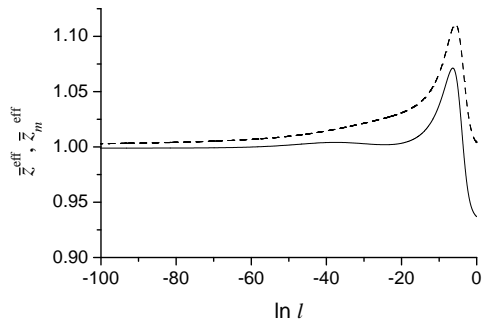}}}
\caption{\label{fig6} Dependencies of normalized effective dynamical
critical exponents of the order parameter and conserved density
correspondent to flow 4. Notations as if Fig. \ref{fig5}.}
\end{figure}

The effective behavior of the two dynamical critical exponents for
the order parameter and the conserved density might be quite
different as one sees comparing Figs. \ref{fig2} and \ref{fig3}.
However, one may ask if both exponents reach the asymptotic values
in the same way. For this purpose we introduce a normalization of
the values of the effective exponents by their values in the
asymptotics. In particular, we introduce notations
$\overline{z}^{\rm eff}={z}^{\rm eff}/z$, $\overline{z}_{m}^{\rm
eff}={z}_{m}^{\rm eff}/z_{m}$ for order parameter exponent and
conserved density exponent correspondingly. Figs.~\ref{fig5},
~\ref{fig6} show behavior of normalized exponents for order
parameter and conserved density for flows 2 and 4 correspondingly.
It illustrates that approach to the asymptotics for order parameter
exponents and conserved density exponents occurs in different way
for different flows, that means for different initial conditions.
For system with small degree of disorder (small $u(0)$ and $w(0)$,
flow 4) the approach of order parameter dynamical exponent to
asymptotic regime is  faster than for the conserved density one,
while for system with larger amount of disorder (flow 2) approach of
both quantities is almost simultaneous.

\section{Conclusion\label{V}}

In this paper, we have studied model C dynamics of the random
anisotropy magnets with cubic distribution of local anisotropy axis.
For this purpose two-loop dynamical RG function $\zeta_{\Gamma}$ has
been obtained. On the base of static results\cite{Dudka05} the
dependencies of effective critical exponents of order parameter,
$z^{\rm eff}$, and conserved density, $z_{m}^{\rm eff}$, on the flow
parameter were calculated.

The two-loop approximation adopted in our paper may be considered as
certain compromise between what is feasible in static calculations
from one side, and in dynamic ones form the other side. As a matter
of fact, the state-of-the-art expansions of the static RG functions
in the minimal subtraction scheme are currently available for many
models within the five-loop accuracy \cite{review,Kleinert} but it
is  not the case for the dynamic functions. Complexity of dynamical
calculations is reflected in the current situation, when the results
beyond two loops have been obtained for model A only. The model C
even with no structural disorder seems to be outside present
manageable problems (see the recent review \cite{Folk06}). However,
there are examples which demonstrate the even in two loops highly
accurate results for dynamical characteristics can be obtained. One
of them is given by the critical dynamics of $^4$He at the
superfluid phase transition \cite{helium}. Besides, analysis of the
two-loop static RG functions refined by resummation also brings
about sufficiently accurate quantitative characteristics of a static
critical behavior in disordered systems \cite{review,Jug83}.

In the asymptotics the conserved density is decoupled from the order
parameter and the dynamical critical behavior of random anisotropy
model with cubic random axis distribution is the same as that of the
random-site Ising model. Crossover occurring between different FPs
present in the random anisotropy model considerably influences the
non-asymptotic critical properties. Different scenarios of dynamical
critical behavior  are observed depending of the initial values of
the model parameters. The main feature is the presence of additional
peaks on the curves for the effective dynamical critical exponents
in comparison with the effective model A critical dynamics.

As far as the  approach to the asymptotics is very slow, the
effective exponents may be observed in experiments and in numerical
simulations. The effective exponent for the order parameter may take
a value far away from the asymptotic one (the asymptotic value in in
our two loop calculation is $z=2.139$). The same holds for the
conserved density effective critical exponent which may be far of
its van Hove asymptotic value $z_{m}=2$.  For example one can
observe values of $z^{\rm eff}$ and $z_{m}^{\rm eff}$ close to those
for pure Ising model with model C dynamics.

This work was supported by Fonds zur F\"orderung der
wissenschaftlichen Forschung under Project No. P16574

\appendix
\section{\label{A} Perturbation expansion}

\begin{figure}[htbp]
\begin{picture}(500,70)
\put(5,45){\includegraphics[width=0.1\textwidth]{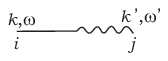}}\put(85,50)
{$G(k,\omega)\delta(k{+}k')\delta(\omega{+}\omega')\delta_{i,j}$}
\put(5,15){\includegraphics[width=0.1\textwidth]{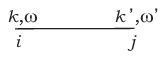}}\put(85,20)
{$C(k,\omega)\delta(k{+}k')\delta(\omega{+}\omega')\delta_{i,j}$}
\put(230,50){\includegraphics[width=0.1\textwidth]{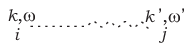}}\put(300,50)
{$H(k,\omega)\delta(k{+}k')\delta(\omega{+}\omega')\delta_{i,j}$}
\put(230,20){\includegraphics[width=0.1\textwidth]{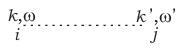}}\put(300,20)
{$D(k,\omega)\delta(k{+}k')\delta(\omega{+}\omega')\delta_{i,j}$}
\end{picture}
\caption{\label{fig7} Propagators for  constructing Feynman graphs.
$G(k,\omega)$ and $H(k,\omega)$ are response propagators while
$C(k,\omega)$ and $D(k,\omega)$ are correlation propagators.}
\end{figure}

We perform our calculations on the basis of the Lagrangian defined
by (\ref{rLagrangian}) using the Feynman graph technique. The
propagators for this Lagrangian are shown in the Fig.~\ref{fig7}.

\begin{figure}[htbp] \vspace{0.5cm}
\begin{picture}(500,160)
\put(40,140){a}
\put(60,120){\includegraphics[width=0.15\textwidth]{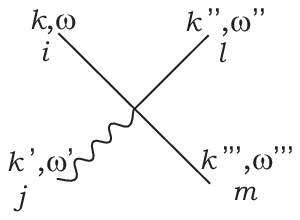}}\put(0,100)
{$\Gamma
A\delta(k+k'+k''+k''')\delta(\omega+\omega'+\omega''+\omega''') $
}\put(40,60){b}
\put(60,40){\includegraphics[width=0.17\textwidth]{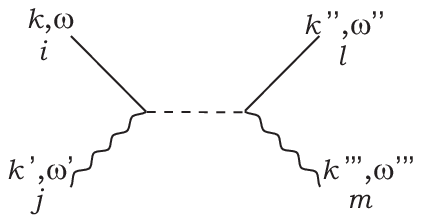}}\put(0,20)
{$\Gamma^2
B\delta(k+k')\delta(k''+k''')\delta(\omega+\omega')\delta(\omega''+\omega''')
$ } \put(270,140){c}
\put(290,120){\includegraphics[width=0.15\textwidth]{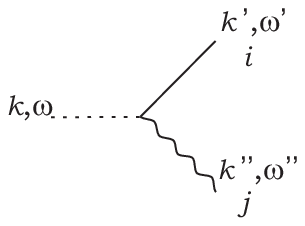}}\put(250,100)
{$\mathring{\Gamma}\mathring{\gamma}\delta(k+k'+k'')\delta(\omega{+}\omega'{+}\omega'')
\delta_{i,j}$ } \put(270,60){d}
\put(290,40){\includegraphics[width=0.15\textwidth]{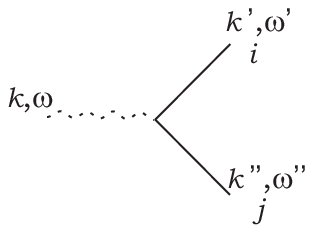}}\put(250,20)
{$\frac{1}{2}\mathring{\lambda}\mathring{\gamma}
k^2\delta(k+k'+k'')\delta(\omega{+}\omega'{+}\omega'') \delta_{i,j}$
}
\end{picture}
\caption{\label{fig8} Vertices for our  model. In vertex {\bf a},
$A$ stands for ${v_0}/{3!}\,
{(\delta_{i,j}\delta_{l,m}+\delta_{i,l}\delta_{j,m}+\delta_{i,m}\delta_{j,l})}/{3}$
or ${y_0}/{3!}\,\delta_{i,j}\delta_{j,l}\delta_{l,m}$. In vertex
{\bf b},  $B$ stands for ${u_0}/{3!}\,\delta_{i,j}\delta_{l,m}$ or
${w_0}/{3!}\,\delta_{i,j}\delta_{j,l}\delta_{l,m}$. Vertices {\bf c}
and {\bf d} originate from the coupling to the conserved density.}
\end{figure}

Response propagators $G(k,\omega)$ and $H(k,\omega)$ are equal to
\begin{equation} \hspace{-2em}
G(k,\omega)=1/(-i\omega+\mathring{\Gamma}(\mathring{\tilde{\mu}}+k^2))
\qquad\mbox{and}\qquad H(k,\omega)=1/(-i\omega+\mathring{\lambda}
k^2)  \,  ,
\end{equation}
 while the correlation propagators $C(k,\omega)$ and $D(k,\omega)$ are equal to
\begin{equation}\hspace{-3em}
C(k,\omega)=2\mathring{\Gamma}/|-i\omega+\mathring{\Gamma}(\mathring{\tilde{\mu}}+k^2)|^2
\qquad\mbox{and}\qquad D(k,\omega)=2\mathring{\lambda}
k^2/|-i\omega+\mathring{\lambda}k^2|^2 \, .
\end{equation}
The vertices defined by  Lagrangian are shown in Fig.~\ref{fig8}.

We obtain an expression for the two-point vertex function
${{\mathring{\Gamma}_{\tilde{\varphi}\varphi}}}^{i,j}$ by keeping
the diagrams up to two-loop order. The result of calculations can be
expressed in form:
\begin{eqnarray}\label{form}
 \mathring{\Gamma}_{\tilde{\varphi}\varphi}(\xi,k,\omega)=
-i\omega\mathring{\Omega}_{\tilde{\varphi}\varphi}(\xi, k,\omega)+
\mathring{\Gamma}^{st}_{{\varphi}\varphi}(\xi, k)\mathring{\Gamma}
\, .
 \end{eqnarray}
Here we introduce the correlation length
$\xi(\mathring\mu=\mathring{\tilde \mu}+ {\mathring{\gamma}
\mathring{h}},\mathring{u},\mathring{v},\mathring{w},\mathring{y})$,
which is defined by
\begin{eqnarray}
\xi^2=\left.\frac{\partial\ln\mathring{\Gamma}^{st}_{\varphi\varphi}}{\partial
k^2 }\right|_{k^2=0} \, .
 \end{eqnarray}
The function $\mathring{\Gamma}_{{\varphi}\varphi}$ is the static
two-loop vertex function of the disordered magnet. The structure
(\ref{form}) of the dynamic vertex function of pure model C was
obtained in Ref. \cite{Folk02} up to two-loop order.

We can express two-loop dynamical function
$\mathring{\Omega}_{\tilde{\varphi}\varphi}$ in the following form:
\begin{eqnarray}\label{om}
 \mathring{\Omega}_{\tilde{\varphi}\varphi}(\xi,k,\omega)=1+
\mathring{\Omega}^1_{\tilde{\varphi}\varphi}(\xi,k,\omega)+
\mathring{\Omega}^2_{\tilde{\varphi}\varphi}(\xi,k,\omega),
 \end{eqnarray}
where the one loop contribution reads:
\begin{eqnarray}\label{om1} \hspace{-3em}
\mathring{\Omega}^1_{\tilde{\varphi}\varphi}(\xi,k,\omega)=-\frac{\mathring{u}+\mathring{w}}{3}\mathring{\Gamma}\int_{k'}
\frac{1}{(-i{\omega}+\mathring{\Gamma}(\xi^{-2}+k'^2))(\xi^{-2}{+}k'^2)}+
{\gamma}\mathring{\Gamma}I_C(\xi,k,\omega),
 \end{eqnarray}
while the two-loop contribution is of the form:
\begin{eqnarray}\label{om2}
 \mathring{\Omega}^2_{\tilde{\varphi}\varphi}(\xi,k,\omega)=
\mathring\Gamma(\frac{n+2}{18} {\mathring v}^2+\frac{{\mathring
y}^2}{6}+\frac{{\mathring v}{\mathring y}}{3}){\mathring
W}^{(A)}_{{\tilde\varphi}\varphi}(\xi,k,\omega)- \nonumber\\
\mathring\Gamma(\frac{n+2}{3} {\mathring v}+{\mathring
y}){{\mathring\gamma}^2}{\mathring
C}^{(T3)}_{{\tilde\varphi}\varphi}(\xi,k,\omega)+\mathring\Gamma
{{\mathring\gamma}^4}{\mathring
S}_{{\tilde\varphi}\varphi}(\xi,k,\omega)+\nonumber\\\mathring\Gamma(\frac{n+2}{9}
{\mathring v}{\mathring u}+\frac{{\mathring y}{\mathring w}}{3}
+\frac{{\mathring w}{\mathring v}}{3}+\frac{{\mathring y}{\mathring
u}}{3}){\mathring
W}^{(CD2)}_{{\tilde\varphi}\varphi}(\xi,k,\omega)+\mathring\Gamma
(\frac{{\mathring u}^2}{9}+\frac{{\mathring
w}^2}{9}+2\frac{{\mathring w}{\mathring u}}{9})\times
\nonumber\\
\left({\mathring
W}^{(CD3)}_{{\tilde\varphi}\varphi}(\xi,k,\omega)+{\mathring
W}^{(CD4)}_{{\tilde\varphi}\varphi}(\xi,k,\omega)\right)-\nonumber\\
\mathring\Gamma
(\frac{\mathring{u}+\mathring{w}}{3}){\mathring\gamma}^2\left({\mathring
W}^{(CD5)}_{{\tilde\varphi}\varphi}(\xi,k,\omega)+{\mathring
W}^{(CD6)}_{{\tilde\varphi}\varphi}(\xi,k,\omega)+2{\mathring
W}^{(CD7)}_{{\tilde\varphi}\varphi}(\xi,k,\omega)\right).
 \end{eqnarray}
 %with rescaled coupling
 %$\mathring{\gamma}=\mathring{\gamma}_m/{\sqrt {a_m}} $.

In (\ref{om1}) and  (\ref{om2})  the expressions for the integrals
$I_C$, ${\mathring W}^{(A)}$, ${\mathring C}^{(T3)}$ and ${\mathring
S}$ of the pure model C are given in the Appendix A.1 in Ref.
\cite{Folk04}, while the contributions for ${\mathring W}^{(CDi)}$
are presented in the Appendix of Ref. \cite{Dudka06b}.

Following the renormalization procedure  for
$\mathring{\Gamma}_{\tilde\varphi\varphi}$ we obtain the two-loop
renormalizating factor $Z_{\tilde \varphi}$:
\begin{eqnarray}\label{zphitilde}
 Z_{\tilde
\varphi}=1{+}\frac{2}{\varepsilon}\frac{u+w}{3}-2\frac{\gamma^2}{\varepsilon}\frac{W}{1+W}
+\frac{1}{\varepsilon^2}\Big[\left({\gamma^2}\frac{W}{1+W}\left(\frac{1}{1+W}
-\left(\frac{n}{2}-1\right)\right)- \right . \nonumber\\ \left .
 \left(\frac{n+2}{3}v+y\right)\right){\gamma^2}\frac{W}{1+W}-
\left(5+\frac{W}{1{+}W}\right)\frac{u+w}{3}{\gamma^2}\frac{W}{1{+}W}\!+\frac{2}{3}u^2+
\nonumber\\
\frac{4}{3}uw+\frac{2}{3}w^2+\frac{n+2}{9}uv+\frac{yu}{3}+
\frac{vw}{3}+\frac{yw}{3}\Big]{+} \nonumber\\
\frac{1}{2\varepsilon}\Bigg\{\!\Bigg[\left(\frac{n{+}2}{3}v+y\right)\left(\!1{-}3\ln\frac{4}{3}\!\right){+}
{\gamma^2}\!\frac{W}{1{+}W}\Bigg(\!\frac{n}{2}{-}
\frac{W}{1{+}W}{-}\frac{3(n{+}2)}{2}\ln\frac{4}{3}{-}
\nonumber\\
\frac{1{+}2W}{1{+}W}\ln\frac{(1{+}W)^2}{1{+}2W}\!\Bigg)\Bigg]{\gamma^2}\frac{W}{1{+}
W}{-}
%\nonumber\\
\frac{11}{36}u^2{-}\frac{3(n+2)}{36}uv{-}\frac{11}{18}uw{-}\frac{uy}{4}{-}
\nonumber\\
\frac{1}{4}vw{-}({u+w}){\gamma^2}\frac{W}{1{+}W}\Bigg\}{-}
\frac{1}{\varepsilon}\left(\frac{n{+}2}{12}{v^2}+\frac{vy}{6}+\frac{y^2}{4}\right)\left(\ln\frac{4}{3}{-}\frac{1}{12}\right){-}
\nonumber\\
\frac{1}{\varepsilon}\frac{u+w}{3}{\gamma^2}\frac{W}{1+W}\left[\frac{W}{2}\ln\frac{W}{1+W}-
 \frac{3}{2}\ln(1+W) -\frac{1}{2}\frac{W}{1+W}\ln W\right].
 \end{eqnarray}

\end{document}